\documentclass{article}
\usepackage[paper=letterpaper,margin=2cm]{geometry}
\usepackage{soul, color, xcolor}
\usepackage{graphicx}
\usepackage{amsmath}
\usepackage{amssymb}
\usepackage{braket}
\usepackage{bm}
\usepackage{amsfonts}
\usepackage{cite}
\usepackage[ruled,boxed,linesnumbered]{algorithm2e}
\usepackage{tcolorbox}
\usepackage[colorlinks=true,citecolor=blue]{hyperref}
\usepackage{booktabs}
\usepackage{dashrule}
\usepackage{authblk}
\usepackage{orcidlink}
\usepackage{multirow}
\tcbuselibrary{breakable}
\tcbset{colbacktitle=yellow!10!white, colback=white, breakable, coltitle=black, before upper={\parindent10pt\noindent}, fonttitle=\bfseries}
\date{}
\title{\textbf{Experimental Measurement-Device-Independent Quantum Key distribution with Flawed State-Preparation over 300 km}}

\author{\begin{minipage}{0.92\textwidth}
    \centering
    \small
    Yi-Fei Lu\orcidlink{0000-0001-8212-0556}$^1$, Yan-Yang Zhou$^1$, Yang Wang\orcidlink{0000-0001-9774-1064}$^1$, Yu Zhou$^1$, Xiao-Lei Jiang$^1$, Xin-Hang Li$^2$, Hai-Tao Wang$^1$, Yan-Mei Zhao$^1$, Jia-Ji Li$^1$, Chun Zhou$^1$, Hong-Wei Li$^1$, Lin-Jie Zhou$^{2,\dagger}$, Wan-Su Bao$^{1,*}$

    $^1$Henan Key Laboratory of Quantum Information and Cryptography, IEU, Zhengzhou 450001, China

    $^2$State Key Laboratory of Advanced Optical Communication Systems and Networks,
    
    Shanghai Key Lab of Navigation and Location Services, Department of Electronic Engineering,
    
    Shanghai Jiao Tong University, Shanghai 200240, China

    $^\dagger$ljzhou@sjtu.edu.cn, $^*$bws@qiclab.cn

\end{minipage}}

\begin{document}
\maketitle

{\centering\section*{Abstract}}

Quantum key distribution (QKD) promises theoretically secure communication. However, it encounters challenges in implementation security and performance due to inevitable device imperfections. Since the proposal of measurement-device-independent (MDI) QKD, the critical step toward practical security is to secure QKD with imperfect sources. The source imperfections manifest as state-preparation uncertainty (SPU) in various aspects, e.g., encoding uncertainty, intensity fluctuation, and imperfect vacuum states. Here, we perform an MDI-QKD experiment and achieve both high practical security and superior performance. We address the general form of SPU and guarantee a tight estimation of the secret key rate based on the operator dominance method. We achieve secure key distribution over 303.37 km, which not only represents the farthest distance in experiments involving SPU but also considers the most SPU scenarios. Our experimental results represent a significant step toward promoting practical and secure quantum communication.

\

{\bf Keywords}: Quantum key distribution; State-preparation uncertainty; Measurement-device-inde-pendent

\clearpage

\section{Introduction}

Quantum key distribution (QKD) \cite{bennett2014RN153, xu2020RN130, pirandola2020RN485, portmann2022RN813} enables two remote parties, Alice and Bob, to achieve information-theoretically secure communication. Despite its promise, the practical application of QKD faces various theoretical and experimental challenges. The disparity between idealized theoretical models and real-world QKD systems may compromise practical security and limit performance metrics such as the secret key rate (SKR) and transmission distance. While measurement-device-independent (MDI) QKD \cite{lo2012RN72,braunstein2012RN581} has significantly mitigated quantum attacks on the receiver, addressing imperfections at the source remains an urgent task to ensure robust security and optimal performance.

Many breakthrough experiments in MDI-QKD have been demonstrated, significantly improving key aspects such as SKR, transmission distance, repetition frequency, and integration \cite{tang2015RN1286,yin2016RN288,comandar2016RN1287,liu2019RN530,bacco2019RN1290,wei2020RN1030,semenenko2020RN1171,cao2020RN1199,lowndes2021RN1289}. These advancements highlight the potential for efficient long-distance quantum communication networks. However, strict assumptions are required on the source side that the quantum states are prepared without flaws. Hence, a crucial step to enhance practical security is securing MDI-QKD with imperfect sources. The role of the source is to prepare pure states according to predefined settings. In the widely employed decoy-state QKD \cite{hwang2003RN202, lo2005RN81, wang2005RN185}, this involves preparing weak coherent states with precise intensity and encoding parameters, e.g., polarization \cite{liao2017RN965, wei2020RN969, li2023RN987}, phase \cite{lucamarini2013RN961, bernd2017RN962, yuan2018RN513}, or time-bin encoding \cite{boaron2018RN286, henry2020RN678, fan2022RN1150, fadri2023RN1023}. However, the practical prepared states may be uncertain due to limitations in modulation precision and environmental disturbances. The experimental results indicate that the state-preparation uncertainties (SPUs) of both encoding and intensity follow the Gaussian distribution \cite{huang2023RN959}. Addressing this discrepancy is crucial for enhancing both the practical security and performance of MDI-QKD systems.

Several security proofs have been developed to address SPU in QKD systems. The Gottesman-Lo-Lütkenhaus-Preskill (GLLP) security analysis \cite{gottesman2004RN266} could accommodate these issues through fidelity analysis, but the provable SKR decreases rapidly as channel loss increases. Some proofs provide conservative analysis based on simple characterizations of SPU intervals \cite{nagamatsu2016RN64,mizutani2019RN61,huang2023RN959}. On the one hand, certain security proofs offer security analysis of QKD with only encoding SPU. For instance, the loss-tolerant (LT) method can achieve a high SKR by characterizing the encoding SPU \cite{tamaki2014RN69,xu2015RN402,mizutani2015RN435,nagamatsu2016RN64,tang2016RN433,bourassa2020RN648}. Alternatively, the virtual mutually unbiased bases method \cite{li2024RN1179} can be employed to address the characterized encoding imperfections. It also could address the encoding SPU by regarding the encoding SPU as uncharacterized qubit sources \cite{yin2014RN1161,wang2016RN1159,hwang2017RN1162,zhu2019RN1158}. Besides, the issue of unbalanced basis misalignment, a specific case of encoding SPU, has been analyzed with tolerant analysis in MDI-QKD \cite{lu2022RN835}. On the other hand, current methods to address intensity SPU (i.e., intensity fluctuations) involve estimating the counting rates of single-photon states by considering the worst-case scenario using analytical formulas or linear programming \cite{wang2007RN1145, wang2009RN84, li2014RN733, jiang2022RN1146}. Additionally, the issue of imperfect vacuum states is particularly critical due to the finite extinction ratio. However, the vacuum state is essential and serves as the signal state in the time-bin encoding scheme. The issues of the imperfect vacuum states have been analyzed in twin-field QKD \cite{hu2022RN1152} and side-channel-free QKD \cite{jiang2022RN746}. 

At present, some MDI-QKD experiments have been performed to address only the encoding SPU \cite{tang2016RN433,wang2019RN1154,zhou2020RN494,lu2022RN835}. In this work, to obtain both high performance and enhanced practical security, we perform an MDI-QKD experiment by considering the general form of SPU, including encoding uncertainty, intensity fluctuation, and the imperfect vacuum state. The core idea of our method is to construct operator inequalities, known as operator dominance method \cite{maeda2019RN507}, to provide a tight estimation of the amount of secure keys. The operator dominance method was proposed in Ref.  \cite{maeda2019RN507} to simplify the security proof of twin-field QKD in the ﬁnite-size regime. It could estimate the counting rate of one state by simply constructing the operator inequality with test states. Here, we employ this method to address the SPUs, leveraging its simplicity to accommodate various types of SPUs within a unified theoretical framework. To validate the proposed framework, we have successfully achieved a 303.37 km MDI-QKD over standard single-mode optical fiber in our experiment. Compared with previous MDI-QKD experiments which only consider the encoding SPU with a maximum distance of 170 km \cite{zhou2020RN494}, our results not only address more imperfect issues but also greatly extend the distance by 83.56 km in the finite case. Furthermore, compared to long-distance experiments that do not consider SPUs \cite{yin2016RN288}, our system maintains a similar secret key rate even at approximately 250 km. This shows that our approach performs robustly over long distances, paving the way for future advancements in secure quantum communication.

\section{Results}

\subsection{Protocol}
\label{protocol}

Step 1 (State preparation and characterization). In every round, Alice first selects a random bit 0 or 1, and determines the Z or X basis with probabilities $p_z$ or $p_x$, respectively. In Z basis, Alice selects a intensity pair $(\tau_{a_1},\tau_{a_2})$ which will be $(\mu,o)$, $(\nu,o)$, $(o,o)$, $(\mu,\omega)$, $(\nu,\omega)$, $(\omega,\omega)$, $(\mu,\mu)$ or $(\nu,\nu)$ with intensity $\mu > \nu > \omega > o \gtrsim 0$ and the random phases $\theta, \theta' \in [0,2\pi)$. Here the state $o$ is supposed to be 0 under the ideal condition. Then she prepares the coherent states $\ket{e^{i\theta}\sqrt{\tau_{a_1}}} \ket{e^{i\theta'}\sqrt{\tau_{a_2}}}$ or $\ket{e^{i\theta'}\sqrt{\tau_{a_2}}} \ket{e^{i\theta}\sqrt{\tau_{a_1}}}$ when the random bit is 0 or 1. In X basis, Alice selects a intensity $\tau_a \in \{\nu, \omega, o\}$ and the random phase $\theta\in[0,2\pi)$, then prepare the coherent states $\ket{e^{i\theta}\sqrt{\tau_a}} \ket{e^{i\theta}\sqrt{\tau_a}}$ or $\ket{e^{i\theta}\sqrt{\tau_a}} \ket{e^{i(\theta+\pi)}\sqrt{\tau_a}}$ when the random bit is 0 or 1. In practical systems, the intensity and phase may be mixed. Alice could characterize the actual prepared states. In Z basis, as the phases are both random in two bins, the state when the random bit is 0 can be expressed as
\begin{equation}
    \label{eqsc1}
    \rho_{\tau_{a_1}\tau_{a_2}}^z = \sum_{i,j=0}^\infty q_{i,\tau_{a_1}} q_{j,\tau_{a_2}} \ket{ij} \bra{ij}.
\end{equation}
The state $\rho_{\tau_{a_2}\tau_{a_1}}^z$ when the random bit is 1 can be defined similarly. Here, we have defined that $q_{i,\tau} = \int_{x \in \Pi_{\tau}} f_x p_{i,x} dx$, where $\Pi_{\tau}$ represents the set of intensity distribution corresponding to the setting $\tau$, $f_x$ denotes the probability density function, and $p_{i,x} = e^{-x} x^i/i!$ is the Poisson distribution probability. In X basis, the relative phase between two bins is fixed as 0 or $\pi$ in the ideal case, but the actual states with encoding SPU can be defined as $\rho_{\tau_a}^{x_0}$ and $\rho_{\tau_a}^{x_1}$ when the random bit is 0 and 1 with
\begin{equation}
    \label{eqsc3}
    \rho_{\tau_a}^{x_{0(1)}} = \sum_{k=0}^{\infty} q_{k,\tau_a} \int_{y\in\Lambda_{0(1)}} g_y \ket{\chi_{k,y}}\bra{\chi_{k,y}} dy \triangleq \sum_{k=0}^{\infty} q_{k,\tau_a} \rho^{x_{0(1)}}_k,
\end{equation}
where $\Lambda_{0(1)}$ represents the set of phase distribution and $g_y$ denotes the probability density function, and the two-mode $k$-photon state is defined as
\begin{equation}
    \ket{\chi_{k,y}} = \frac{1}{\sqrt{2^k}} \sum_{r=0}^k \sqrt{C_k^r} e^{iry} \ket{r,k-r}.
\end{equation}
Bob independently selects the parameters and prepares the states in the same way. The prepared states are sent to Charlie through the quantum channel. Here, we note that the prepared states in Eqs. (\ref{eqsc1}) and (\ref{eqsc3}) implicitly imply that the states are independent and identically distributed (i.i.d.) and do not include the potential side-channels. These are summarized as the i.i.d. assumption and the qubit assumption, enabling us to confine our analysis to the modulation spaces.

Step 2 (Measurement). Charlie is supposed to perform the interference measurement and announce the measurement results $\{0,1,\perp\}$. Here, 0 and 1 corresponds to a measurement of the Bell state $\ket{\psi_{0}} = (\ket{01}+\ket{10})/\sqrt{2}$ and $(\ket{01}-\ket{10})/\sqrt{2}$, and $\perp$ corresponds to other failed measurements. The honesty of Charlie is not necessary and the deception will affect the SKR but will not cause security issues.

Step 3 (Sifting). Alice and Bob sift the effective rounds when the measurement results are 0 or 1, and announce the selected intensity and basis for those rounds. They generate the bits for those effective rounds when both are Z basis and the intensity pairs are $(\mu,o)$ or $(o,\mu)$. They keep these bits as raw bits while Bob flips all of his bits. Other effective rounds are used to estimate the parameters.

Step 4 (Parameter estimation and postprocessing). They estimate the counting rate $s_{11}$ and phase error rate $e_x$ when both sides sent the single-photon states. The detailed estimation methods can be found in Sec. \ref{sec_method}. By applying the error correction and privacy amplification methods, they distill the secret key bits from the raw bits, and the SKR is shown as \cite{lo2012RN72,ma2012RN73}
\begin{equation}
    R_\text{MDI} \geq p_z^2 p_\mu^2 \{q_{(1,1)} s_{11} [1 - h(e_x)] - f Q_{[\mu,\mu]} h(E_z) \},
\end{equation}
where $p_z$ is the probability of selecting the Z basis and $p_\mu$ is the probability of choosing the intensity pairs $(\mu,o)$ or $(o,\mu)$, $q_{(1,1)} = q_{1,\mu}^2 q_{0,o}^2$, and $Q_{[\mu,\mu]}$ and $E_z$ are the counting rate and error rate when both sides choosing the intensity pairs $(\mu,o)$ or $(o,\mu)$.

\subsection{Experimental setup}

\begin{figure*}
    \centering
    \includegraphics[scale = 0.4]{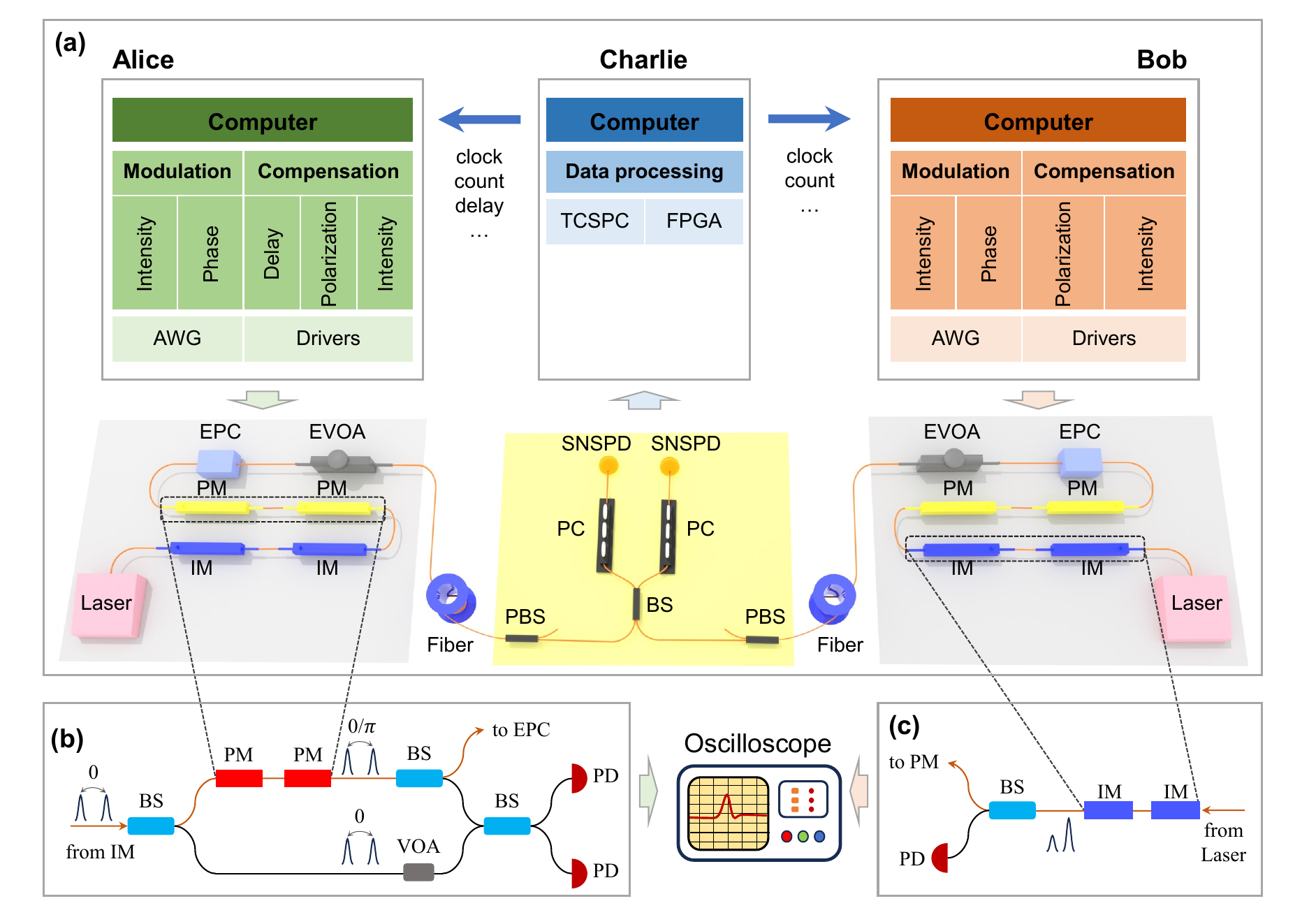}
    \caption{Experimental set-up. (a) MDI-QKD experimental system. The MDI-QKD system is controlled by three computers located at Alice, Bob, and Charlie. At the senders (Alice and Bob), the AWG and hardware drivers are employed to provide modulation signals while enabling system tracking, compensation, and precise calibration. At the receiver (Charlie), it performs interference measurement, samples the detection results, and provides feedback control information. The entire system is capable of continuous and stable operation. Within the system's optical path, the continuous-wave light is chopped into a train of pulses with four intensities by two IMs. Then the phases are modulated by two PMs for phase encoding and phase randomization. These four modulators are driven by an AWG, and the biases of the IMs are controlled by a bias controller. The polarization and time delay are compensated according to the feedback from Charlie. The encoded pulses are attenuated to the single-photon level by an EVOA and transmitted to Charlie. Bob's setup is the same as Alice's, but their modulations are independent. Alice, Bob, and Charlie are connected with the standard single-mode fiber. Charlie aligns the pulses from Alice and Bob with two PBSs and performs interference with a 50:50 BS. The interference pulses are detected with two SNSPDs. The electrical output signals of SNSPDs are processed with TCSPC and FPGA. The counting rate within 300 milliseconds of a SNSPD is sent to Alice and Bob for polarization compensation. The delay between Alice and Bob is sent to Alice for delay compensation. (b) Phase characterization scheme. The pulse is first split into two pulses, then one arm is modulated by PMs and the other arm serves as the reference pulses without phase modulation. After phase modulation, the signal pulse is then split with one for QKD and the other for characterization. The delay and attenuation of the two arms are balanced with fiber and VOA. The phase difference between two arms can be calculated according to the intensity ratio of two PDs. Then the modulated phase difference between the two bins is obtained. (c) Intensity characterization scheme. The signal pulses without attenuation are directly split with one detected with a PD and the other for QKD. AWG, arbitrary waveform generator; IM, intensity modulator; PM, phase modulator; EPC, electrically polarization controller; EVOA, electrically variable optical attenuator; PBS, polarization beam splitter; BS, beam splitter; PC, polarization controller; SNSPD, superconducting nanowire single-photon detector; TCSPC, time-correlated single-photon counting system; FPGA, field programmable gate array; PD, photodetector.}
    \label{fig_mdi_setup}
\end{figure*}

As illustrated in Fig. \ref{fig_mdi_setup}(a), the MDI-QKD experimental setup consists of two senders, Alice and Bob, and a measurement node, Charlie. Each sender employs a free-running continuous-wave (CW) laser with a central wavelength of 1550.12 nm. The two lasers are independent without frequency locking. For this setup, we choose time-bin encoding as the Z basis and phase encoding as the X basis. A pulse train is modulated by two cascaded intensity modulators (IMs) with a temporal width of 400 ps and a repetition rate of 500 MHz. The first IM chops the CW light into fixed-intensity or vacuum pulses. Then the second IM modulates the fixed-intensity pulses into the signal or decoy states and enhances the extinction ratio of the vacuum pulses. In this way, we could generate four different intensity levels of optical pulses, and the intensity pairs fulfill the requirements of step 1 in Sec. \ref{protocol}. Note that the intensity pairs in Z and X bases are different. Two phase modulators (PMs) then randomize the phase of the pulses and modulate the phase difference between the two time bins. When the Z basis is selected, the two pulses in a pair are independently modulated with random phases. When the X basis is chosen, the first pulse in a pair is modulated with a random phase $\theta$, while the second pulse is modulated with either $\theta$ or $\theta + \pi$, depending on the random bit. The modulators are driven by an arbitrary waveform generator and the IMs are biased with bias controllers. Two electrically polarization controllers (EPCs) compensate for polarization drift based on feedback from Charlie. Finally, the light pulses are attenuated to the single-photon level using electrically variable optical attenuators (EVOAs).

The quantum channel is the standard single-mode fiber with a typical loss coefficient of 0.2 dB/km. At the receiver, the light pulses first pass through polarization beam splitters (PBSs) to ensure identical polarization. Then the pulses interfere at the 50:50 beam splitter (BS). After adjusting the polarization with polarization controllers (PCs), the interference pulses are detected using superconducting nanowire single-photon detectors (SNSPDs). Considering that the period of the light pulses is 2 ns while the dead time of the SNSPDs is around 50 ns, Charlie will only announce the detection results of the Bell state $\ket{\psi_1} = (\ket{01} - \ket{10}) / \sqrt{2}$ or the failed detection $\perp$. The detection efficiency and dark count rate of the SNSPDs are calibrated to approximately $55\%$ and 3E$-8$ in the sample interval. The detection signals are sampled and processed with a time-correlated single-photon counting system (TCSPC) and a field programmable gate array (FPGA).

Charlie provides feedback to Alice and Bob on the counting rate of the one SNSPD every 300 milliseconds. Alice and Bob alternately adjust the EPCs to maximize the detector counting rate for polarization compensation, thereby achieving polarization tracking and compensation. Besides, Charlie provides feedback on the delay of Alice and Bob's pulses, and Alice adjusts the delay to achieve alignment. Alice and Bob both synchronize their operations with Charlie.

Alice and Bob characterize the SPUs with setups in Figs. \ref{fig_mdi_setup}(b) and (c). The phase characterization setup is shown in \ref{fig_mdi_setup}(b). In MDI-QKD, the two PMs will modulate the phase of two time bins, with the encoding based on the phase difference between them. To characterize the SPU, we split the pulses into two arms: one for phase modulation and the other serving as a reference. Then the split pulses are interfered with at a 50:50 beam splitter (BS). The pulses after interference are detected with classical photodetectors (PDs). We could calculate the phase difference of the interfered pulses. The difference in the phase differences calculated from the two bins represents the modulated phase. The intensity characterization setup is shown in \ref{fig_mdi_setup}(c), where a PD is directly used to measure pulses of varying intensities.

In the above steps, it can be seen that the entire QKD process can be divided into two phases: the key distribution mode and the SPU characterization mode. The switching between the two modes can be efficiently implemented using the optical switch or BS. As shown in the Figs. \ref{fig_mdi_setup}(b) and (c), we present a BS-based implementation for the two operational modes, enabling simultaneous source characterization and key generation. The brown lines represent the link for key generation, which incorporates additional BSs. The only cost of this configuration is the introduction of additional attenuation at the source side, which will not affect the key generation and can be compensated by adjusting the EVOA at the output. The characterization of SPUs can be flexibly conducted at any point during the key distribution process, allowing for precise and thorough assessment without compromising efficiency.

\subsection{Experimental results}

We implement the experiment with the standard single-mode fiber at the transmission distances of 101.15, 152.40, 202.31, 253.56, and 303.57 km. The frequency difference between Alice and Bob's lasers is kept within 4 MHz without frequency locking, which is shown in Fig. \ref{fig_laser_fre_diff}. We note that the frequency drift is caused only by the internal operation of the lasers. This means that the frequency drift is independent from the QKD setting choices and hence the impact of frequency drift is uniform across different encoded states. Besides, as the phase randomization is additionally performed in the state preparation step, there are no phase correlations between different pulses. Therefore, the i.i.d. assumption can be guaranteed in the presence of frequency drift.

\begin{figure*}
    \centering
    \includegraphics[scale = 0.55]{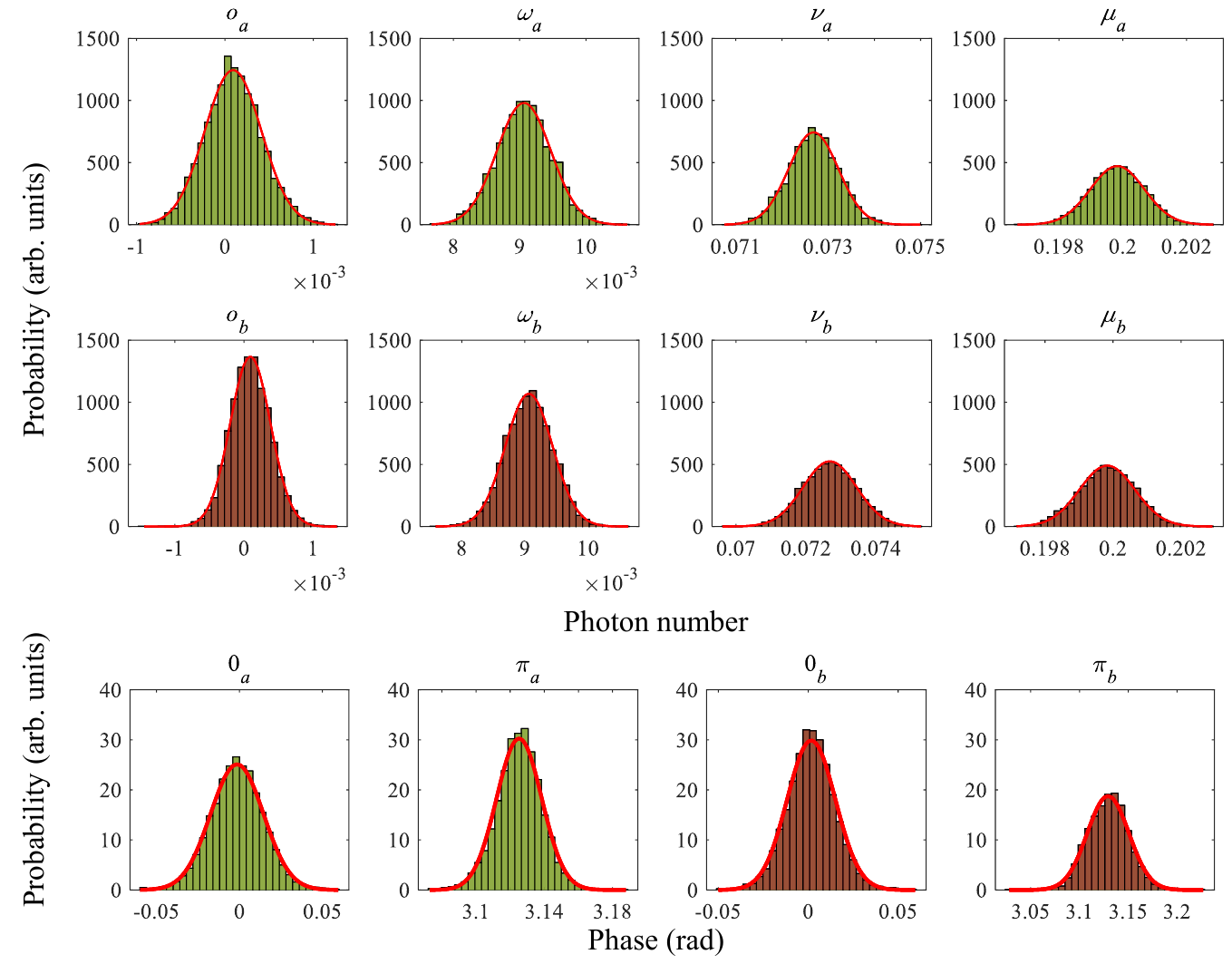}
    \caption{Results of intensity and phase characterization. The first two rows illustrate the intensity distribution, where intensity $\mu$ represents the signal state, while the others denote the decoy states. The last row shows the phase distribution. The subscripts $a$ and $b$ indicate Alice and Bob, respectively.}
    \label{fig_phase_intensity}
\end{figure*}

Before the experiments at every distance, we characterize the modulated phase and intensity. The results show that the phase and intensity of the light pulses obey the Gaussian distribution. The characterized intensity SPU in the 303.57 km experiment and the characterized phase SPU are shown in Fig. \ref{fig_phase_intensity}. The detailed results are listed in Tabs. \ref{tab_mdi_phase} and \ref{tab_mdi_intensity}.

\begin{figure}
    \centering
    \includegraphics[scale = 0.55]{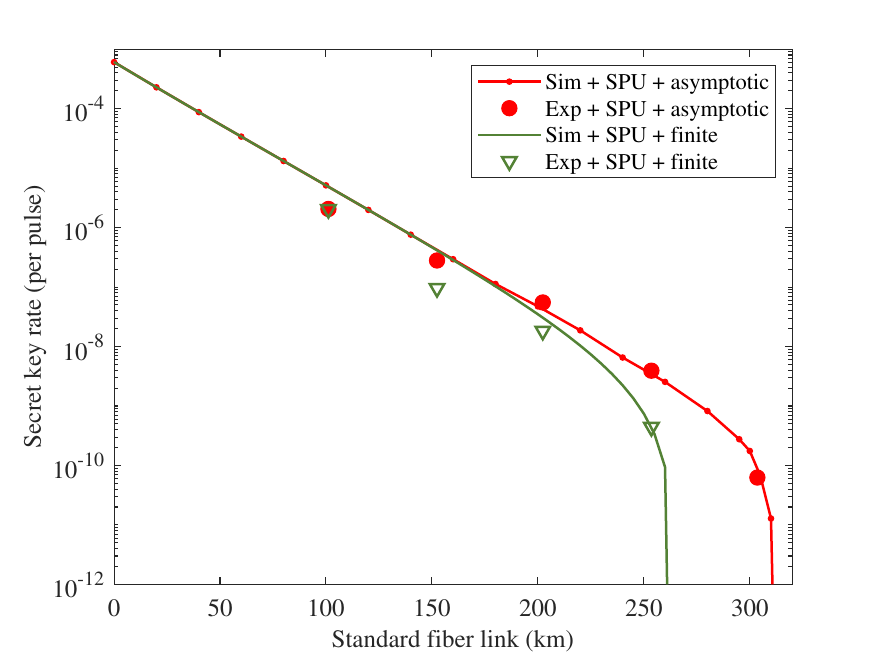}
    \caption{Secret key rate (SKR) in logarithmic scale versus transmission distance between Alice and Bob. The SKR is simulated according to the characterized results in our experiments in the asymptotic and finite cases. The experimental SKR results are shown with red dots and green triangles in the asymptotic and finite cases, respectively.}
    \label{fig_mdi_results}
\end{figure}

The simulated and experimental SKR results are shown in Fig. \ref{fig_mdi_results}. The SKR is simulated based on the characterized results at 253.56 km in both the asymptotic and finite cases. And the simulation method is given in App. \ref{simulation}. The experimental SKR results are represented by red dots and green triangles for the asymptotic and finite cases, respectively. The results show that a secure key distribution distance of 303.57 km can still be achieved even when simultaneously accounting for the encoding SPU, intensity modulation SPU, and the effects of non-ideal vacuum states. More experimental details are shown in App. \ref{detailed_results}. In addition, we consider the finite effects \cite{curty2014RN456} with security coefficient $\varepsilon = 1$E$-10$, and perform experiments with the total number $N=$ 7.68E12, 7.50E13, 9.67E13 and 5.08E14 at 101.15, 152.40, 202.31, 253.56 km, respectively. In the finite case, the achievable distance remains at 253.56 km. This distance is a significant improvement over previous experiments, demonstrating the robustness and reliability of our system under realistic conditions.

We compare our results with recent MDI-QKD experiments that consider SPUs in Tab. \ref{tab_exp_result_com}. Current experiments primarily focus on encoding SPU while neglecting the issues related to intensity modulation and non-ideal vacuum states. Our work accounts for the largest variety of SPUs while achieving the longest secure key distribution distance. At present, the farthest distance achieved in MDI-QKD experiments considering only encoding SPU is 170 km. We have significantly extended this distance to 303.37 km, demonstrating a substantial improvement in transmission capability. Moreover, our system achieves higher secret key rates. For instance, at a distance of 152.40 km, the rate reaches 23.9 bps, which surpasses the rate previously reported for a 50 km distance. Furthermore, when compared to long-distance experiments that do not account for SPUs \cite{yin2016RN288}, our system achieves a rate that remains on the same order of magnitude even at approximately 250 km. This indicates that our approach maintains robust performance and efficiency over extended distances, highlighting its superiority in practical long-distance quantum communication scenarios. 

\begin{table*}
    \caption{List of recent MDI-QKD experiments with SPUs. Three types of SPUs are considered: the encoding SPU, the intensity modulation SPU, and the effects of imperfect vacuum states.}
    \label{tab_exp_result_com}
    \centering
    \resizebox{\textwidth}{!}{
    \begin{tabular}{ccccccccccc}
    \toprule
    Reference                                   & Encoding                      & Intensity                     & Vacuum                        & Distance (km)          & Rate (bps)  & Year                  & Notes \\ \midrule
    \multirow{2}{*}{Yin \cite{yin2016RN288}}    & \multirow{2}{*}{$\times$}     & \multirow{2}{*}{$\times$}     & \multirow{2}{*}{$\times$}     & 259                    & 0.221       & \multirow{2}{*}{2016} & \multirow{2}{*}{Without SPU} \\ 
                                                &                               &                               &                               & 404$^{\text a}$        & 3.21E$-$04  &                       &   \\ \midrule
    Tang \cite{tang2016RN433}                   & $\checkmark$                  & $\times$                      & $\times$                      & 40$^{\text b}$         & 10          & 2016                  & / \\ \midrule
    Wang  \cite{wang2019RN1154}                 & $\checkmark$                  & $\times$                      & $\times$                      & 50                     & 18.8        & 2019                  & / \\ \midrule
    Zhou \cite{zhou2020RN494}                   & $\checkmark$                  & $\times$                      & $\times$                      & 170                    & 3.53        & 2020                  & / \\ \midrule
    Wei \cite{wei2020RN1030}                    & $\times$                      & $\times$                      & $\times$                      & 180                    & 31          & 2020                  & Without SPU  \\ \midrule
    Lu \cite{lu2022RN835}                       & $\checkmark$                  & $\times$                      & $\times$                      & 50                     & 21.4        & 2022                  & / \\ \midrule
    \multirow{5}{*}{This work}                  & \multirow{5}{*}{$\checkmark$} & \multirow{5}{*}{$\checkmark$} & \multirow{5}{*}{$\checkmark$} & 101.15                 & 510         & \multirow{5}{*}{2025} & \multirow{5}{*}{/} \\
                                                &                               &                               &                               & 152.40                 & 23.9        &                       &   \\
                                                &                               &                               &                               & 202.31                 & 4.58        &                       &   \\
                                                &                               &                               &                               & 253.56                 & 0.11        &                       &   \\
                                                &                               &                               &                               & 303.37$^{\text b}$     & 0.0157      &                       &   \\ \bottomrule
    \end{tabular}}\\
    \footnotesize{$^{\text a}$Ultra-low-loss fiber.}
    \footnotesize{$^{\text b}$Asymptotic.}
\end{table*}

\section{Methods}
\label{sec_method}

Following the security analysis framework \cite{koashi2007RN398,koashi2009RN137,tamaki2014RN69}, a key step is to estimate the counting rate and error rate of the single-photon states. To facilitate this analysis, we consider the virtual entanglement version of the MDI-QKD protocol in the presence of SPU. In both the prepare-and-measure and entanglement versions, the classical and quantum information available to Eve, and the ultimate results for Alice and Bob are consistent. In the entanglement version, the phase error rate can be defined to quantify the information leakage to Eve. Therefore, the security is equivalent and the security of the QKD protocol can be guaranteed by proving the virtual entanglement version. When utilizing the weak coherent source in MDI-QKD systems, it is crucial to estimate the single-photon counting rate under various conditions to obtain the counting rate and error rate of the single-photon states. In this scenario, the SPU occurs in both the encoding space and the intensity space. Especially, the vacuum states are essential for time-bin encoding but are impractical due to the limited extinction ratio. To tightly bound the observations of single-photon states in the presence of encoding and intensity SPUs, we construct operator dominance conditions, as summarized in Eqs. (\ref{eqsc8}) and (\ref{eqsc5}). This could relax the stringent requirements on state preparation while ensuring the robustness and performance of QKD systems.

\subsection{Estimation of the phase error rate}

To analyze the phase error rate, we consider the single-photon MDI-QKD where both Alice and Bob prepare the single-photon state and send them to Charlie. The single-photon states in Z basis are $\rho_1^{z_0} = \ket{01}$ or $\rho_1^{z_1} = \ket{10}$, a fraction in Eq. (\ref{eqsc1}). The encoding states in these forms are guaranteed with the qubit assumptions, such that the states in high dimension is identical. We note that the qubit assumption can be removed by decomposing the non-qubit states into a direct sum of two states with one in a qubit space and the other one in the complementary space \cite{pereira2019RN65,gnanapandithan2025RN1303}, enabling a more general practical security analysis. In the entanglement protocol, Alice prepares the general state in Z basis as
\begin{equation}
    \ket{\phi}_{ac_1} = \sum_{i\in\{0,1\}} \sqrt{p_i} \ket{i}_a \ket{\bar{i},i}_{c_1},
\end{equation}
where $p_0$ and $p_1$ are the probabilities of random bits 0 and 1, and are equal to 0.5 ideally. Here the ancillary system $a$ is kept secretly and the actual system $c_1$ is sent to Charlie. Similarly, Bob prepares the states $\ket{\phi}_{bc_2}$ and sends system $c_2$ to Charlie. 

Then the untrusted Charlie is supposed to perform the interference measurement and announce the results. The channel evolution and Charlie's operation on the composite systems $c_1 c_2$ can be characterized by the POVM $\{\hat{C}_0, \hat{C}_1, \hat{C}_\perp\}$, corresponding to the announcement of the measurement results $\{0,1,\perp\}$. Note that there is no assumption about POVM due to the feature of measurement device independent. 

The phase error rate corresponding to the measurement results 0 and 1 can be shown as
\begin{equation}
    \begin{aligned}
        \label{eqsc2}
        e_{x_0} = \frac{\text{Tr} [(\ket{+-}\bra{+-} + \ket{-+}\bra{-+}) \otimes \hat{C}_0 \hat{P}(\ket{\phi}_{ac_1} \otimes \ket{\phi}_{bc_2})]}{\text{Tr} [\hat{C}_0 \hat{P}(\ket{\phi}_{ac_1} \otimes \ket{\phi}_{bc_2})]}, \\
        e_{x_1} = \frac{\text{Tr} [(\ket{++}\bra{++} + \ket{--}\bra{--}) \otimes \hat{C}_1 \hat{P}(\ket{\phi}_{ac_1} \otimes \ket{\phi}_{bc_2})]}{\text{Tr} [\hat{C}_1 \hat{P}(\ket{\phi}_{ac_1} \otimes \ket{\phi}_{bc_2})]},
    \end{aligned}
\end{equation}
where $\hat{P}(\ket{\cdot}) = \ket{\cdot}\bra{\cdot}$. The overall phase error rate can be defined based on the convexity of the binary
entropy function as
\begin{equation}
    e_x = \frac{\sum_{i} e_{x_i} \text{Tr} [\hat{C}_i (\rho_{c_1} \otimes \rho_{c_2})]}{\text{Tr} [(\mathbb{I} - \hat{C}_\perp) (\rho_{c_1} \otimes \rho_{c_2})]}.
\end{equation}
Note that due to the dead-time of SPDs, the announcement of the Bell state $\ket{\psi_0}$ may be impossible and $e_{x_1}$ is just the final phase error rate in most MDI-QKD systems. The denominators in Eq. (\ref{eqsc2}) are just the counting rate of announcing 0 and 1 when Alice and Bob prepare the composite states $\ket{\phi}_{ac_1} \otimes \ket{\phi}_{bc_2}$. The elements in the numerators in Eq. (\ref{eqsc2}) are shown as
\begin{equation}
    \label{eqsc4}
    \text{Tr} [\ket{jk}\bra{jk} \otimes \hat{C}_i \hat{P}(\ket{\phi}_{ac_1} \otimes \ket{\phi}_{bc_2})] = \frac{1}{4} \text{Tr} [ \hat{C}_i (\varrho_j \otimes \varrho_k) ],
\end{equation}
where $i\in\{0,1\}$, $j,k\in\{+,-\}$, and $\varrho_\pm = \hat{P}[(\ket{01} \pm \ket{10}) / \sqrt{2}]$. 

To estimate the phase error rate, we analyze the states in X basis. The single-photon states in X basis are $\rho_1^{x_0}$ and $\rho_1^{x_1}$, a fraction of the states in Eq. (\ref{eqsc3}). To obtain the upper bound of the phase error rate, we give the operator dominance condition as
\begin{equation}
    \begin{aligned}
        \label{eqsc8}
        \varrho_\pm \leq \sigma_\pm \triangleq \sum_{d \in \{z_0,z_1,x_0,x_1\}} \beta_{d}^\pm \rho_1^{d},
    \end{aligned}
\end{equation}
with $\beta_{d} \in \mathbb{R}$. The details for the contribution of Eq. (\ref{eqsc8}) are shown in App. \ref{construction}. Then we can obtain that
\begin{equation}
    \varrho_j \otimes \varrho_k \leq \varrho_j \otimes \sigma_k \leq \sigma_j \otimes \sigma_k, j,k\in\{+,-\}.
\end{equation}
This can be used to estimate the upper bound of Eq. (\ref{eqsc4}) as
\begin{equation}
    \begin{aligned}
        \label{eqsc6}
    \text{Tr} [ \hat{C}_i (\varrho_j \otimes \varrho_k) ] &\leq \text{Tr} [ \hat{C}_i (\sigma_j \otimes \sigma_k) ] = \sum_{d,d^\prime\in\{z_0,z_1,x_0,x_1\}} \beta_{d}^{j} \beta_{d^\prime}^k \text{Tr} [ \hat{C}_i ( \rho_1^{d} \otimes \rho_1^{d^\prime} )],
    \end{aligned}
\end{equation}
where $\text{Tr} [ \hat{C}_i ( \rho_1^{z_i} \otimes \rho_1^{z_{i^\prime}} )]$ is the probability of announcing measurement result $i$ when Alice and Bob prepare the states $\rho_1^{d}$ and $\rho_1^{d^\prime}$, respectively. As these probabilities can be obtained directly in the single-photon MDI-QKD, the phase error rate can be estimated in the presence of encoding SPU. In the coherent source MDI-QKD, the phase error rate can be estimated in the same way on condition that the bounds of the probability $\text{Tr} [ \hat{C}_i ( \rho_1^{z_i} \otimes \rho_1^{z_{i^\prime}} )]$ are estimated with the decoy-state method. We present the analysis in the next subsection.

\subsection{Estimation of the single-photon counting rate}

In the coherent source MDI-QKD, the actual prepared states are shown in Eqs. (\ref{eqsc1}) and (\ref{eqsc3}). The decoy-state method \cite{hwang2003RN202,lo2005RN81,wang2005RN185} can be applied to estimate different counting rates of the single-photon states. Below we show how to estimate the bounds of the single-photon counting rate to calculate the lower bound of $s_{11}$ and the upper bound of the phase error rate $e_x$. As we consider the SPUs of the encoding, intensity modulation, and imperfect vacuum states, the decoy-state estimation should be modified.

To estimate these parameters, we give the operator dominance conditions as
\begin{equation}
    \begin{aligned}
        \label{eqsc5}
        \rho_1^d & \geq \sum_{(\tau_1,\tau_2)} \beta_{\tau_1,\tau_2}^{d,0} \rho_{\tau_1,\tau_2}^z, d \in \{z_0,z_1\},\\
        \rho_1^d & \leq \sum_{(\tau_1,\tau_2)} \beta_{\tau_1,\tau_2}^{d,1} \rho_{\tau_1,\tau_2}^z, d \in \{z_0,z_1\},\\
        \rho_1^d & \geq \sum_{\tau} \beta_{\tau}^{d,0} \rho_{\tau}^x, d \in \{x_0,x_1\},\\
        \rho_1^d & \geq \sum_{\tau} \beta_{\tau}^{d,1} \rho_{\tau}^x, d \in \{x_0,x_1\},
    \end{aligned}
\end{equation}
with proper $\beta_{\tau_1,\tau_2}^{d,0}, \beta_{\tau_1,\tau_2}^{d,1}, \beta_{\tau}^{d,0}, \beta_{\tau}^{d,1} \in \mathbb{R}$. The details for the contribution of Eq. (\ref{eqsc5}) are shown in App. \ref{construction}. The validity of Eq. (\ref{eqsc5}) relies on the absence of correlations, such as those introduced by the pattern effect \cite{yoshino2018RN71}. When considering the pattern effect, we could utilize the pattern sifting and alternate key distillation post-processing operations to remove the impact of correlations \cite{yoshino2018RN71}. Besides, we also could utilize reference technique \cite{pereira2020RN255} to accommodate the correlations between setting choices, such as bit and basis choices. Here, we consider that the i.i.d. assumption is guaranteed.

With the operator dominance conditions in Eq. (\ref{eqsc5}), we show how to estimate the parameters. The counting rate of the single-photon state $s_{11}$ is defined as
\begin{equation}
    s_{11} = \frac{1}{4} \sum_{d,d^\prime\in\{z_0,z_1\}}\text{Tr}[(\hat{C}_0 + \hat{C}_1) (\rho_1^d \otimes \rho_1^{d^\prime})].
\end{equation}
According to the first operator dominance condition in Eq. (\ref{eqsc5}), we can conclude the following formula with $d,d^\prime \in \{z_0,z_1\}$
\begin{equation}
    \label{eqsc7}
    \rho_1^d \otimes \rho_1^{d^\prime} \geq \sum_{(\tau_1,\tau_2)} \sum_{(\tau_3,\tau_4)} \beta_{\tau_1,\tau_2}^{d,0} \beta_{\tau_3,\tau_4}^{d^\prime,0} \rho_{\tau_1,\tau_2}^z \otimes \rho_{\tau_3,\tau_4}^z.
\end{equation}
Then the lower bound of the single-photon counting rate $s_{11}$ is given by
\begin{equation}
    \begin{aligned}
        s_{11} \geq & \frac{1}{4} \sum_{d,d^\prime\in\{z_0,z_1\}} \sum_{(\tau_1,\tau_2)} \sum_{(\tau_3,\tau_4)} \beta_{\tau_1,\tau_2}^{d,0} \beta_{\tau_3,\tau_4}^{d^\prime,0} \times \text{Tr}[(\hat{C}_0 + \hat{C}_1) (\rho_{\tau_1,\tau_2}^z \otimes \rho_{\tau_3,\tau_4}^z)],
    \end{aligned}
\end{equation}
where $\text{Tr}[(\hat{C}_0 + \hat{C}_1) (\rho_{\tau_1,\tau_2}^z \otimes \rho_{\tau_3,\tau_4}^z)]$ is the counting rate of announcing 0 and 1 when Alice and Bob prepare the states $\rho_{\tau_1,\tau_2}^z$ and $\rho_{\tau_3,\tau_4}^z$, respectively.

According to Eqs. (\ref{eqsc2})-(\ref{eqsc6}), to obtain the upper bound of the phase error rate $e_x$, we only need to estimate the lower or upper bounds of $\text{Tr} [ \hat{C}_i ( \rho_1^{d} \otimes \rho_1^{d^\prime})]$ with $d,d^\prime \in \{z_0,z_1,x_0,x_1\}$ according to the sign of the coefficient $\beta_d \beta_{d^\prime}$ in Eq. (\ref{eqsc6}). The bounds can be calculated the same as Eq. (\ref{eqsc7}). We directly give the results in the following. For $d,d^\prime \in \{z_0,z_1\}$, we have
\begin{equation}
    \begin{aligned}
        \label{eqsc11}
        \text{Tr} [ \hat{C}_i ( \rho_1^{d} \otimes \rho_1^{d^\prime})] \geq \sum_{(\tau_1,\tau_2)} \sum_{(\tau_3,\tau_4)} \beta_{\tau_1,\tau_2}^{d,0} \beta_{\tau_3,\tau_4}^{d^\prime,0} \text{Tr}[\hat{C}_i (\rho_{\tau_1,\tau_2}^z \otimes \rho_{\tau_3,\tau_4}^z)],\\
        \text{Tr} [ \hat{C}_i ( \rho_1^{d} \otimes \rho_1^{d^\prime})] \leq \sum_{(\tau_1,\tau_2)} \sum_{(\tau_3,\tau_4)} \beta_{\tau_1,\tau_2}^{d,1} \beta_{\tau_3,\tau_4}^{d^\prime,1} \text{Tr}[\hat{C}_i (\rho_{\tau_1,\tau_2}^z \otimes \rho_{\tau_3,\tau_4}^z)].
    \end{aligned}
\end{equation}
For $d \in \{z_0,z_1\}$ and $d^\prime \in \{x_0,x_1\}$, we have
\begin{equation}
    \begin{aligned}
        \label{eqsc12}
        \text{Tr} [ \hat{C}_i ( \rho_1^{d} \otimes \rho_1^{d^\prime})] \geq \sum_{(\tau_1,\tau_2)} \sum_{\tau} \beta_{\tau_1,\tau_2}^{d,0} \beta_{\tau}^{d^\prime,0} \text{Tr}[\hat{C}_i (\rho_{\tau_1,\tau_2}^z \otimes \rho_{\tau}^x)],\\
        \text{Tr} [ \hat{C}_i ( \rho_1^{d} \otimes \rho_1^{d^\prime})] \leq \sum_{(\tau_1,\tau_2)} \sum_{\tau} \beta_{\tau_1,\tau_2}^{d,1} \beta_{\tau}^{d^\prime,1} \text{Tr}[\hat{C}_i (\rho_{\tau_1,\tau_2}^z \otimes \rho_{\tau}^x)].
    \end{aligned}
\end{equation}
For $d \in \{x_0,x_1\}$ and $d^\prime \in \{z_0,z_1\}$, we have
\begin{equation}
    \begin{aligned}
        \label{eqsc13}
        \text{Tr} [ \hat{C}_i ( \rho_1^{d} \otimes \rho_1^{d^\prime})] \geq \sum_{\tau} \sum_{(\tau_3,\tau_4)} \beta_{\tau}^{d,0} \beta_{\tau_3,\tau_4}^{d^\prime,0} \text{Tr}[\hat{C}_i (\rho_{\tau}^x \otimes \rho_{\tau_3,\tau_4}^z)],\\
        \text{Tr} [ \hat{C}_i ( \rho_1^{d} \otimes \rho_1^{d^\prime})] \leq \sum_{\tau} \sum_{(\tau_3,\tau_4)} \beta_{\tau}^{d,1} \beta_{\tau_3,\tau_4}^{d^\prime,1} \text{Tr}[\hat{C}_i (\rho_{\tau}^x \otimes \rho_{\tau_3,\tau_4}^z)].
    \end{aligned}
\end{equation}
And for $d,d^\prime \in \{x_0,x_1\}$, we have
\begin{equation}
    \begin{aligned}
        \label{eqsc14}
        \text{Tr} [ \hat{C}_i ( \rho_1^{d} \otimes \rho_1^{d^\prime})] \geq \sum_{\tau} \sum_{\tau^\prime} \beta_{\tau}^{d,0} \beta_{\tau^\prime}^{d^\prime,0} \text{Tr}[\hat{C}_i (\rho_{\tau}^x \otimes \rho_{\tau^\prime}^x)],\\
        \text{Tr} [ \hat{C}_i ( \rho_1^{d} \otimes \rho_1^{d^\prime})] \leq \sum_{\tau} \sum_{\tau^\prime} \beta_{\tau}^{d,1} \beta_{\tau^\prime}^{d^\prime,1} \text{Tr}[\hat{C}_i (\rho_{\tau}^x \otimes \rho_{\tau^\prime}^x)].
    \end{aligned}
\end{equation}
These parameters $\text{Tr}[\hat{C}_i (\rho_{\tau_1,\tau_2}^z \otimes \rho_{\tau_3,\tau_4}^z)]$, $\text{Tr}[\hat{C}_i (\rho_{\tau_1,\tau_2}^z \otimes \rho_{\tau}^x)]$, $\text{Tr}[\hat{C}_i (\rho_{\tau}^x \otimes \rho_{\tau_3,\tau_4}^z)]$ and $\text{Tr}[\hat{C}_i (\rho_{\tau}^x \otimes \rho_{\tau^\prime}^x)]$ can be obtained directly in MDI-QKD systems, which are the probabilities of announcing result $i$ when Alice and Bob prepare the states $\rho_{\tau_1,\tau_2}^z \otimes \rho_{\tau_3,\tau_4}^z$, $\rho_{\tau_1,\tau_2}^z \otimes \rho_{\tau}^x$, $\rho_{\tau}^x \otimes \rho_{\tau_3,\tau_4}^z$ and $\rho_{\tau}^x \otimes \rho_{\tau^\prime}^x$, respectively.

\section{Discussion}
\label{discussion}

In this work, we have significantly advanced the MDI-QKD experiment by addressing the SPUs in various aspects and extending the maximum transmission distance to 303.37 km. To address these practical issues, we construct the operator dominance method and give a tight estimation of the amount of secure keys. While considering more kinds of SPUs, an improvement of 83.56 km has been achieved over previous MDI-QKD experiments that only considered encoding SPU. Additionally, even at the extended distance, our system maintains a rate comparable to long-distance MDI-QKD experiments that do not account for SPUs. These results highlight the practicality and robustness of our approach in practical MDI-QKD systems. Our work improves the practicality of current MDI-QKD systems and provides a foundation for future development and deployment in secure quantum communication.

As we aim to provide a comprehensive solution to the specific class of practical source imperfection only in the modulation spaces, there remain several crucial research directions in other aspects for future work. First, the encoding may be correlated with other degrees of freedom and hence introduces side channels, which might violate the qubit assumption \cite{jiang2012RN180,huang2018RN303,pereira2019RN65,pereira2023RN952,gnanapandithan2025RN1303}. Second, the correlations may also occur between nearby pulses, e.g., the pattern effect \cite{yoshino2018RN71,2024RN1234}, the setting choice correlations \cite{pereira2020RN255,pereira2025RN1330}, and the phase correlations \cite{marcomini2025RN1329}. This will undermine the condition that the quantum states are i.i.d. At present, a number of approaches can be employed to remove these assumptions \cite{yoshino2018RN71,pereira2019RN65,pereira2020RN255,pereira2023RN952,gnanapandithan2025RN1303,pereira2025RN1330}. Therefore, extending our framework to incorporate these methods and account for such imperfections will be an essential direction for achieving higher practical security in QKD for future research.

\section*{Funding}

This work was supported by National Key Research and Development Program of China (Grant No. 2020YFA0309702), Henan Science and Technology Major Project of the Department of Science \& Technology of Henan Province (No. 241100210400), National Natural Science Foundation of China (Nos. U2130205, 62101597), and Natural Science Foundation of Henan (No. 242300421219).

\appendix

\section{Detailed experimental results}
\label{detailed_results}

In this section, we present the detailed experimental results. The Tabs. \ref{tab_mdi_phase} and \ref{tab_mdi_intensity} give the parameters of Gaussian approximation of the phase and intensity distributions, respectively.  The Tab. \ref{tab_mdi_parameters} lists the probabilities and the Tab. \ref{tab_exp_result} shows the detailed experimental results. 

\begin{figure}[h]
    \centering
    \includegraphics[scale = 0.49]{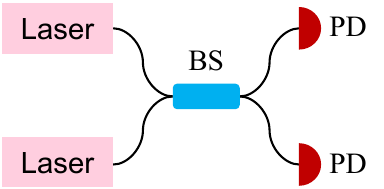}
    \caption{Schematic diagram of the frequency difference characterization scheme. BS, beam splitter; PD, photodetector.}
    \label{fig_beat_setup}
\end{figure}

\begin{figure}[h]
    \centering
    \includegraphics[scale = 0.49]{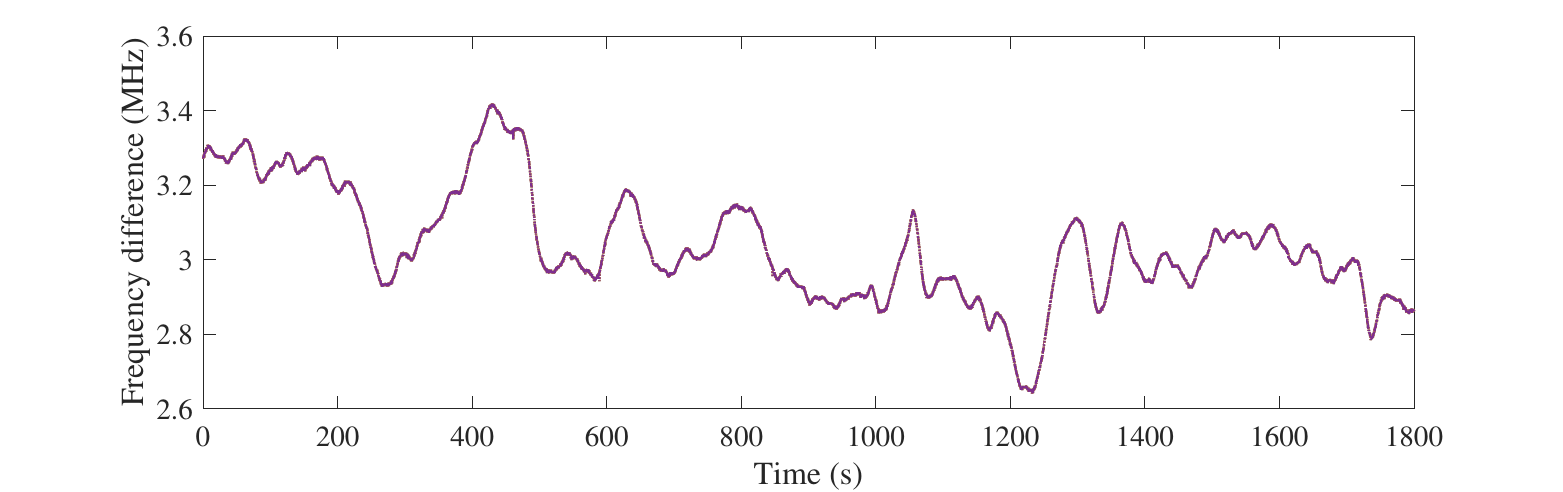}
    \caption{The frequency difference of two free-running lasers.}
    \label{fig_laser_fre_diff}
\end{figure}

Besides, as shown in Fig. \ref{fig_beat_setup}, we measure the frequency difference of two independent lasers through beat frequency without feedback every 0.1 seconds in 30 minutes. The two lasers beat at a 50:50 BS and the results are detected with the classical PDs. The frequency difference then is analyzed with the detected time domain signals. Fig. \ref{fig_laser_fre_diff} shows the frequency difference between two free-running lasers. 

\begin{table}
    \centering
    \caption{Parameters of Gaussian approximation of phase distributions. $\bar{x}$: mean, $\sigma$: standard deviation.}
    \label{tab_mdi_phase}
    \begin{tabular}{ccccc}
    \toprule
              & $0_a$     & $\pi_a$ & $0_b$  & $\pi_b$ \\ \midrule
    $\bar{x}$ & $-$0.0017 & 3.1253  & 0.0015 & 3.1293  \\ 
    $\sigma$  & 0.0172    & 0.0143  & 0.018  & 0.0245  \\ \bottomrule
    \end{tabular}
\end{table}

\begin{table*}
    \centering
    \caption{Parameters of Gaussian approximation of intensity distributions. $\bar{x}$: mean, $\sigma$: standard deviation.}
    \label{tab_mdi_intensity}
    \resizebox{\textwidth}{!}{
    \begin{tabular}{cccccccccc}
    \toprule
    distance                &           & $\mu_a$  & $\nu_a$  & $\omega_a$ & $o_a$     & $\mu_b$  & $\nu_b$  & $\omega_b$ & $o_b$     \\ \midrule
    101.15 & $\bar{x}$ & 0.204089 & 0.060832 & 0.005893   & 9.15E$-$5 & 0.206184 & 0.060095 & 0.005776   & 9.93E$-$5 \\ 
                            & $\sigma$  & 0.001058 & 0.000417 & 0.000307   & 0.000375  & 0.001186 & 0.000737 & 0.000453   & 0.000461  \\ \midrule
    152.40 & $\bar{x}$ & 0.200601 & 0.062914 & 0.006276   & 6.29E$-$5 & 0.204206 & 0.062903 & 0.006111   & 0.000137  \\
                            & $\sigma$  & 0.001027 & 0.000603 & 0.000427   & 0.000439  & 0.001096 & 0.000607 & 0.000449   & 0.000511  \\ \midrule
    202.31 & $\bar{x}$ & 0.205472 & 0.061212 & 0.006039   & 7.98E$-$5 & 0.206309 & 0.061031 & 0.006095   & 7.33E$-$5 \\ 
                            & $\sigma$  & 0.00097  & 0.000585 & 0.000409   & 0.00037   & 0.001148 & 0.0006   & 0.000457   & 0.000453  \\ \midrule
    253.56 & $\bar{x}$ & 0.198356 & 0.063045 & 0.008034   & 5.90E$-$5 & 0.200584 & 0.064894 & 0.007884   & 7.82E$-$5 \\ 
                            & $\sigma$  & 0.000843 & 0.000899 & 0.000203   & 0.000132  & 0.000698 & 0.000611 & 0.000214   & 0.000226  \\ \midrule
    303.57 & $\bar{x}$ & 0.199829 & 0.072682 &	0.009065   & 9.20E$-$5 & 0.186857 & 0.072263 & 0.009337   & 0.000118  \\ 
                            & $\sigma$  & 0.000847 & 0.000537 &	0.000407   & 0.000320  & 0.000763 & 0.000712 & 0.000348   & 0.000272  \\ \bottomrule
    \end{tabular}}
\end{table*}

\begin{table*}
    \centering
    \caption{Experimental parameters of the probability. $p_{(\tau_1,\tau_2)}$ denotes the probability of preparing the states with intensities $\tau_1$ and $\tau_2$, where the order of $\tau_1$ and $\tau_2$ is random.}
    \label{tab_mdi_parameters}
    \begin{tabular}{ccccccccc}
    \toprule
          & $p_{(\mu,o)}$ & $p_{(\nu,o)}$ & $p_{(o,o)}$ & $p_{(\nu,\nu)}$ & $p_{(\mu,\mu)}$ & $p_{(\mu,\omega)}$ & $p_{(\nu,\omega)}$ & $p_{(\omega,\omega)}$ \\ \midrule
    Alice & 0.7574        & 0.0382        & 0.0360      & 0.033           & 0.0284          & 0.0278             & 0.0408             & 0.0384                \\ 
    Bob   & 0.7460        & 0.0394        & 0.0378      & 0.035           & 0.0306          & 0.0326             & 0.0384             & 0.0402                \\ \bottomrule
    \end{tabular}
\end{table*}

\begin{table*}
    \centering
    \caption{Experimental results of MDI-QKD at various quantum link fiber lengths.}
    \label{tab_exp_result}
    \begin{tabular}{cccccc}
    \toprule
    distance(km) & $Q_{[\mu,\mu]}$ & $E_z$  & $s_{11}$   & $e_x$  & $R_{\text{MDI}}$ \\ \midrule
    101.15       & 1.44E$-5$       & 0.0047 & 2.83E$-4$  & 0.0939 & 2.06E$-6$        \\ 
    152.40       & 1.78E$-6$       & 0.0036 & 3.55E$-5$  & 0.0845 & 2.80E$-7$        \\ 
    202.31       & 1.82E$-7$       & 0.0025 & 1.21E$-5$  & 0.1880 & 5.51E$-8$        \\ 
    253.56       & 1.89E$-8$       & 0.0027 & 1.30E$-6$  & 0.2331 & 3.92E$-9$        \\ 
    303.37       & 1.83E$-$9       & 0.0043 & 4.80E$-$8  & 0.2703 & 6.26E$-$11       \\ \bottomrule
    \end{tabular}
\end{table*}

\section{Construction of operator dominance conditions}
\label{construction}

In this section, we analyze how to construct the operator dominance conditions in Eqs. (\ref{eqsc8}) and (\ref{eqsc5}).

For Eq. (\ref{eqsc8}), we could restrict it as equality, i.e., $\varrho_\pm = \sum_{d \in \{z_0,z_1,x_0,x_1\}} \beta_{d} \rho_1^{d}$. Then it is equivalent to the following linear equations
\begin{equation}
    \begin{aligned}
        \frac{1}{2}
        \begin{bmatrix}
            2 & 0 & 1                                 & 1\\
            0 & 0 & \int_{y\in\Gamma_0} g_y e^{iy}dy  & \int_{y\in\Gamma_1} g_y e^{iy}dy\\
            0 & 0 & \int_{y\in\Gamma_0} g_y e^{-iy}dy & \int_{y\in\Gamma_1} g_y e^{-iy}dy\\
            0 & 2 & 1                                 & 1
        \end{bmatrix}
        \begin{bmatrix}
            \beta_{z_0}^\pm \\
            \beta_{z_1}^\pm \\
            \beta_{x_0}^\pm \\
            \beta_{x_1}^\pm
        \end{bmatrix}
        = \frac{1}{2}
        \begin{bmatrix}
            1 \\
            \pm 1 \\
            \pm 1 \\
            1 \\
        \end{bmatrix},
    \end{aligned}
\end{equation}
which is also equivalent to the following real version as
\begin{equation}
    \begin{aligned}
        \begin{bmatrix}
            2 & 0 & 1                                 & 1 \\
            0 & 0 & \int_{y\in\Gamma_0} g_y \cos y dy & \int_{y\in\Gamma_1} g_y \cos y dy\\
            0 & 0 & \int_{y\in\Gamma_0} g_y \sin y dy & \int_{y\in\Gamma_1} g_y \sin y dy\\
            0 & 2 & 1                                 & 1
        \end{bmatrix}
        \begin{bmatrix}
            \beta_{z_0}^\pm \\
            \beta_{z_1}^\pm \\
            \beta_{x_0}^\pm \\
            \beta_{x_1}^\pm
        \end{bmatrix}
        =
        \begin{bmatrix}
            1 \\
            \pm 1 \\
            0 \\
            1 \\
        \end{bmatrix}.
    \end{aligned}
\end{equation}
We could sovle this linear equations to obtain the coefficients $\beta_{d}^\pm$ with $d\in\{z_0,z_1,x_0,x_1\}$ for Eq. (\ref{eqsc8}).

The contributions for Eq. (\ref{eqsc5}) are a little complex due to the imperfect vacuum states. We first define the following two operators
\begin{equation}
    \begin{aligned}
        \label{eqsc10}
        \rho^\mu &\triangleq \frac{\rho_{\mu o}^z}{q_{0,\mu}} - \frac{\rho_{oo}^z}{q_{0,o}} = \sum_{j=1}^\infty \alpha_{j,\mu} \ket{j}\bra{j} \otimes \rho_{o},\\
        \rho^\nu &\triangleq \frac{\rho_{\nu o}^z}{q_{0,\nu}} - \frac{\rho_{oo}^z}{q_{0,o}} = \sum_{j=1}^\infty \alpha_{j,\nu} \ket{j}\bra{j} \otimes \rho_{o}
    \end{aligned}
\end{equation}
where the coefficients $\alpha_{j,\tau} = q_{j,\tau} / q_{0,\tau} - q_{j,o} / q_{0,o}$ and the operator $\rho_{o} = \sum_{i=0}^\infty q_{i,o} \ket{i}\bra{i}$. Then the linear combination of these two operators is given as
\begin{equation}
    \label{eqsup6}
    \alpha_{2,\nu} \rho^\mu - \alpha_{2,\mu} \rho^\nu = \sum_{j=1}^{\infty} \beta_j \ket{j}\bra{j} \otimes \rho_{o},
\end{equation}
where the coefficient is defined as 
\begin{equation}
    \begin{aligned}
        \beta_j = \alpha_{2,\nu} \alpha_{j,\mu} - \alpha_{2,\mu} \alpha_{j,\nu} = \frac{1}{q_{0,\mu} q_{0,\nu} q_{0,o} q_{0,o} 2! j!} \times \int_{\mu,\nu,o,o'} f_\mu f_\nu f_o f_{o'} \frac{G(\mu, \nu, o, o', j)}{e^{\mu+\nu+o+o'}} d\mu d\nu do do'.
    \end{aligned}
\end{equation}
The polarity of $\beta_j$ is determined by the numerators $G(\mu, \nu, o, o', j) \triangleq (\nu^2 - o'^2) (\mu^j - o^j) - (\mu^2 - o^2) (\nu^j - o'^j)$. It is easy to verify that $G(\mu, \nu, o, o', 1) < 0$ and $G(\mu, \nu, o, o', 2) = 0$ with the conditions in Eq. (\ref{eqsup8}). This means that $\beta_1 < 0$ and $\beta_2 = 0$. The numerators of $g(\mu, \nu, o, o', j)$ can be re-expressed by changing the order of summation as $G(\mu, \nu, o, o', j) = (\nu^2 - o'^2) (\mu^j - o^j) - (\mu^2 - o'^2) (\nu^j - o^j)$. Note that
\begin{equation}
    \begin{aligned}
        \frac{\mu^{j+1} - o^{j+1}}{\nu^{j+1} - o^{j+1}} - \frac{\mu^j - o^j}{\nu^j - o^j} & = \frac{(\mu+o)(\nu+o)}{(\nu^{j+1} - o^{j+1})(\nu^j - o^j)} \times \sum_{k=0}^{j-1} o^{j-k-1} (\mu^j \nu^k - \nu^j \mu^k) \geq 0.
    \end{aligned}
\end{equation}
Hence we can prove that
\begin{equation}
    \beta_{j+1} \geq \beta_j \times \min_{\nu,o} \Big\{ \frac{\nu^j - o^j}{\nu^{j-1} - o^{j-1}} \Big\}.
\end{equation}
Based on the mathematical induction method, we can conclude that $\beta_j \geq 0$ when $j \geq 2$. Therefore, based on Eq. (\ref{eqsup6}), the following operator dominance conditions is valid
\begin{equation}
    \label{eqsup7}
    \ket{1}\bra{1} \otimes \rho_o \geq \frac{\alpha_{2,\nu} \rho^\mu - \alpha_{2,\mu} \rho^\nu}{\beta_1} = \frac{\alpha_{2,\nu} \rho^\mu - \alpha_{2,\mu} \rho^\nu}{\alpha_{2,\nu} \alpha_{1,\mu} - \alpha_{2,\mu} \alpha_{1,\nu}}.
\end{equation}
At the same time, it is easy to verify that $\alpha_{j,\nu} > 0$, hence the following operator dominance condition is valid
\begin{equation}
    \label{eqsc9}
    \ket{1}\bra{1} \otimes \rho_o \leq \frac{\rho^\nu}{\alpha_{1,\nu}}.
\end{equation}

If the vacuum state is perfect, hence Eqs. (\ref{eqsup7}) and (\ref{eqsc9}) give the first two operator dominance conditions in Eq. (\ref{eqsc5}). Actually, Eq. (\ref{eqsc9}) is irrelevant of the form of the state $\rho_o$, we could obtain the following condition as
\begin{equation}
    \rho_1^{z_0} \leq \frac{1}{q_{0,o}} \sum_{j=1}^\infty \alpha_{j,\nu} \ket{j}\bra{j} \otimes q_{0,o} \ket{0}\bra{0} \leq \frac{\rho^\nu}{q_{0,o} \alpha_{1,\nu}},
\end{equation}
which gives the second condition in Eq. (\ref{eqsc5}). 

To prove the first condition in Eq. (\ref{eqsc5}), we first give the operator inequalities for the (unnormalized) state $q_{0,o}\ket{10}\bra{10} + q_{1,o}\ket{11}\bra{11}$ according to Eqs. (\ref{eqsup7}) and (\ref{eqsc9}), which are shown as
\begin{equation}
    \begin{aligned}
        \label{eqs12}
        \varUpsilon_{o,0} &\triangleq (q_{0,o} + q_{1,o} - 1) \mathbb{I} + \frac{\alpha_{2,\nu} \rho^\mu - \alpha_{2,\mu} \rho^\nu}{\alpha_{2,\nu} \alpha_{1,\mu} - \alpha_{2,\mu} \alpha_{1,\nu}} \leq q_{0,o}\ket{10}\bra{10} + q_{1,o}\ket{11}\bra{11}\\
        \varUpsilon_{o,1} &\triangleq \sum_{(\tau_0,o)\in\Pi'} \beta'_{\tau_0 o, 1} \kappa_{\tau_0} \otimes \kappa_{o} \geq q_{0,o}\ket{10}\bra{10} + q_{1,o}\ket{11}\bra{11}.
    \end{aligned}
\end{equation}
If we prepare another decoy states with preset intensity $\omega$ to replace the vacuum states, we also could obtain the operator inequalities for the (unnormalized) state $q_{0,o}\ket{10}\bra{10} + q_{1,o}\ket{11}\bra{11}$ similar to Eq. (\ref{eqs12}) as $q_{0,\omega}\ket{10}\bra{10} + q_{1,\omega}\ket{11}\bra{11} \geq \varUpsilon_{\omega,0}$ and $\leq \varUpsilon_{\omega,1}$. Therefore, the following operator dominance condition holds
\begin{equation}
    \label{eqs13}
    \rho_1^{z_0} \geq \frac{q_{1,\omega} \varUpsilon_{o,0} - q_{1,o} \varUpsilon_{\omega,1}}{q_{1,\omega} q_{0,o} - q_{1,o} q_{0,\omega}},
\end{equation}
which gives the first condition in Eq. (\ref{eqsc5}). The reason that the estimation is tight is that $q_{0,o} + q_{1,o} - 1$ is close enough to 0.

As the vacuum state is nonessential for X basis, the last two operator dominance conditions in Eq. (\ref{eqsc5}) can be directly obtained according to Eqs. (\ref{eqsup7}) and (\ref{eqsc9}) as
\begin{equation}
    \begin{aligned}
        \rho_1^{d} &\geq \frac{\alpha_{2,\nu} \sigma^{\mu,d} - \alpha_{2,\mu} \sigma^{\nu,d}}{\alpha_{2,\nu} \alpha_{1,\mu} - \alpha_{2,\mu} \alpha_{1,\nu}},\\
        \rho_1^{d} &\leq \frac{\sigma^{\nu,d}}{\alpha_{1,\nu}},
    \end{aligned}
\end{equation}
where the operators are defined similarly to Eq. (\ref{eqsc10}) as
\begin{equation}
    \begin{aligned}
        \sigma^{\mu,d} &\triangleq \frac{\rho_{\mu}^d}{q_{0,\mu}} - \frac{\rho_{o}^d}{q_{0,o}},\\
        \sigma^{\nu,d} &\triangleq \frac{\rho_{\nu}^d}{q_{0,\nu}} - \frac{\rho_{o}^d}{q_{0,o}}.
    \end{aligned}
\end{equation}

The simple and sufficient but not necessary conditions for the above analysis are that for any $\mu \in \Pi_\mu$, $\nu \in \Pi_\nu$, $o,o' \in \Pi_o$, and $\omega,\omega' \in \Pi_\omega$ the following formulas hold
\begin{equation}
    \begin{aligned}
        \label{eqsup8}
        \mu &> o,\\
        \nu &> o,\\
        \mu + o &> \nu + o',\\
        \mu &> \omega,\\
        \nu &> \omega,\\
        \mu + \omega &> \nu + \omega'.
    \end{aligned}
\end{equation}
In perfect QKD systems, the intensities are fixed satisfying $\mu > \nu > \omega > o$, and the conditions in Eq. (\ref{eqsup8}) are trusted. However, the set $\Pi_\tau (\tau \in \{\mu, \nu, \omega, o\})$ becomes complex due to the intensity fluctuation. At this time, the conditions in Eq. (\ref{eqsup8}) are still reasonable. Even in the case that Eq. (\ref{eqsup8}) can not be satisfied, the invalidity is only due to a small portion of states. That is, considering the Gaussian distribution, the probability of preparing those states is small as the actual intensity deviates from the nominal intensity. And the contribution of those partial states is negligible. It is easy to verify the polarity of $\beta_j$ and $\alpha_{j,\nu}$ based on the mathematical induction method. Considering the worst case that the polarity of $\beta_j$ and $\alpha_{j,\nu}$ do not satisfy the above analysis, we could optimize the intensities in advance.

\section{Simulation method}
\label{simulation}

We give the simulation method for $\text{Tr}[\hat{C}_i (\rho_{\tau_1,\tau_2}^z \otimes \rho_{\tau_3,\tau_4}^z)]$, $\text{Tr}[\hat{C}_i (\rho_{\tau_1,\tau_2}^z \otimes \rho_{\tau}^x)]$, $\text{Tr}[\hat{C}_i (\rho_{\tau}^x \otimes \rho_{\tau_3,\tau_4}^z)]$, and $\text{Tr}[\hat{C}_i (\rho_{\tau}^x \otimes \rho_{\tau^\prime}^x)]$ in Eqs. (\ref{eqsc11}), (\ref{eqsc12}), (\ref{eqsc13}), and (\ref{eqsc14}). As their are two bins, we define a unified definition as $Q_{\tau_{a_1}\tau_{a_2},\tau_{b_1}\tau_{b_2}}^{zz(zx,xz,xx)}$, where $(\tau_{a_1},\tau_{a_2})$ and $(\tau_{b_1},\tau_{b_2})$ are Alice and Bob's intensity choices in two bins, and the superscript denotes the bases. The counting rate can be simulated as
\begin{equation}
    \begin{aligned}
        Q_{\tau_{a_1}\tau_{a_2},\tau_{b_1}\tau_{b_2}}^{zz(zx,xz,xx)} &= \text{Int}_{\tau_{a_1},\tau_{a_2},\tau_{b_1},\tau_{b_2},\theta,\vartheta} \big\{e^{-L(\tau_{a_1},\tau_{b_1},\theta)} (1-p_d) \\
        & \times [1 - e^{-R(\tau_{a_1},\tau_{b_1},\theta)} (1-p_d)] \times [1 - e^{-L(\tau_{a_2},\tau_{b_2},\vartheta)} (1-p_d)] \\
        &+ [1 - e^{-L(\tau_{a_1},\tau_{b_1},\theta)} (1-p_d)] \times e^{-R(\tau_{a_1},\tau_{b_1},\theta)} (1-p_d) \\
        &\times [1 - e^{-R(\tau_{a_2},\tau_{b_2},\vartheta)} (1-p_d)]\big\},
    \end{aligned}
\end{equation}
where $\text{Int}$ denotes the integral and the probability density functions are omitted, and we have defined that
\begin{equation}
    \begin{aligned}
        L(\tau_a,\tau_b,\theta) = \frac{\eta}{2}(\tau_a+\tau_b) + \eta \sqrt{\tau_a\tau_b} \cos\theta,\\
        R(\tau_a,\tau_b,\theta) = \frac{\eta}{2}(\tau_a+\tau_b) - \eta \sqrt{\tau_a\tau_b} \cos\theta.
    \end{aligned}
\end{equation}
Here, $\theta,\vartheta \in [0,2\pi)$ for $zz,zx,xz$. But for $xx$, $\vartheta = \theta + \vartheta_a + \vartheta_b$, where $\vartheta_a,\vartheta_b$ denotes the phase-encoding uncertainty. The counting rate $Q_{[\mu,\mu]}$ and the bit error rate $E_z$ can be simulated as
\begin{equation}
    \begin{aligned}
        Q_{[\mu,\mu]} &= p_0 p_1 Q_{\mu_a o_a, \mu_b o_b}^{zz} + p_0 p_0 Q_{\mu_a o_a, o_b \mu_b}^{zz} + p_1 p_1 Q_{o_a \mu_a, \mu_b o_b}^{zz} + p_1 p_0 Q_{o_a \mu_a, o_b \mu_b}^{zz},\\
        E_z &= \frac{1}{Q_z} (p_0 p_1 Q_{\mu_a o_a, \mu_b o_b}^{zz} + p_1 p_0 Q_{o_a \mu_a, o_b \mu_b}^{zz}).
    \end{aligned}
\end{equation}

\bibliographystyle{spphys}
\bibliography{Ref}

\end{document}